\documentclass[12pt]{iopart}
\usepackage{inputenc}
\usepackage{amssymb}
\usepackage{wrapfig}\usepackage{ifpdf}
\ifpdf
	\usepackage[pdftex]{graphicx}
	\usepackage{epstopdf}	
\else
	\usepackage[dvips]{graphicx}
\fi
\usepackage{epsfig,color}
\usepackage{float}
\usepackage{caption}
\usepackage{subcaption}
\newcommand{\be}{\begin{eqnarray}\label}
\newcommand{\ee}{\end{eqnarray}}
\newcommand{\prt}{\partial}
\newcommand{\p}{\prime}
\newcommand{\bib}{\bibitem}

\begin{document}

\title[Algebraic roots of Newtonian mechanics]{Algebraic roots of Newtonian 
mechanics: correlated dynamics of particles on a unique worldline}

\author{Vladimir V. Kassandrov and Ildus Sh. Khasanov}

\address{Institute of Gravitation and Cosmology, Russian Peoples' Friendship University, Moscow, Russia}

\section{Introduction. Pure algebra as the key to physical interactions?}

Origins of the experimentally observable (and extremely intricated!) structure of 
fundamental interactions, of their laws, intensities and scale dependence 
still look as an enigma. It might seem that cardinal solution to this eternal 
problem is hidden in the geometry of physical space-time. 
However, the Minkowski geometry is too 
``soft'' and allows for a wide variety of relativistic invariant interactions, 
if even the gauge invariance of the scheme is required. As to various  
geometries of {\it extended} space-time, at present they seem quite indefinite 
by themselves and, moreover, do not predetermine in any way a distinguished 
structure of physical dynamics.

That is why, from time to time, one can meet articles dealing with 
the most profound, {\it elementary} notions of physics and reformulations of these 
on the basis of geometry, algebra, number theory, etc. We are aware that 
such attempts had been undertaken, say, by P.A.M. Dirac, A. Eddington and  
J.A. Wheeler.

Particularly, one of the most beautiful and striking ideas 
dealing with the foundations of theoretical physics is the 
Wheeler-Feynman's conjecture on the so-called ``one-electron Universe''. In his 
famous telephone call to R. Feynman~\cite{FeynmanNobel}, J. Wheeler said: 
``Feynman, I know why all electrons have the same charge and the same mass. ... 
Because they are all the same electron!''. In fact, this conjecture 
based on the notion of {\bf a set of particles located on a single worldline} 
easily explains 
the property of {\it identity} of elementary particles of one kind, 
the processes of annihilation/creation of a pair of ``particle-antiparticle'' 
(in which one treats a ``positron'' as an ``electron'' running backwards in time 
~\cite{FeynmanPositrons}) etc.    

In his Nobel lecture~\cite{FeynmanNobel} Feynman, one of creators of QED, 
confessed that his true goal was the establishment of 
correlations of an ensemble of identical (point-like or smeared, to avoid 
field divergences) particles {\it on a single worldline} through their 
along-light-cone interactions and on the base of 
a unique Lagrange function. Unfortunately, the ``one-electron Universe'' 
paradigm had not been fully realized; one of the reasons for this is its failure 
to explain the particle-antiparticle asymmetry; other more essential reasons 
will be revealed below.

Remarkably, this paradigm gains natural development in the framework of {\it complex 
algebrodynamics}~\cite{AD1,AD2,Qanalys}. In this approach one attempts to derive 
both the space-time geometry and principal dynamical equations for fields and 
particles from the properties of an exceptional algebraic structure, a sort of 
{\it space-time algebra}. In contrast to geometries, one possesses quite definite 
and transparent classification of exceptional linear algebras based on the  
famous theorems of G. Frobenius and A. Hurvitz. For consistency with 
Special Relativity and Minkowski geometry, most often as such structure 
had been considered the algebra ${\bf Q}$ of {\it complex quaternions}
~\cite{Qanalys,Mink,YadPhys}. 

Specifically, in the complex extension ${\mathbb C}{\bf M}$ of the space-time  -- 
vector space of ${\bf Q}$ --  
the dynamics, even on a single worldline, becomes quite nontrivial.  
Unlike the case of real Minkowski space-time ${\bf M}$, under any position 
and displacement of an ``observer'', the equation of {\it complex light cone} -- 
direct generalization of the {\it retardation equation} in {\bf M} -- 
always have a constant and, generally, 
great number of roots. These define a correspondent number of {\it copies} of 
one and the same particle detected by the observer at their different positions 
on a single worldline; in~\cite{PIRT_dubl} these copies 
have been named ``duplicons''. 

In the framework of another approach, one considers a single worldline in 
{\it real} $\bf M$ but allows for superluminar velocities of particles 
(tachyons) along it. In this case, the observer also encounters an arbitrary 
number of copies of one and the same tachyon. Possible existence of such 
copies-``images'' had been noticed in~\cite{GinsbBolot} and examined 
in detail in~\cite{Bolot}. Note that, unlike   
the situation with duplicons in ${\mathbb C}{\bf M}$, the number of such 
{\it images} 
is not generally constant: some two of these can appear or disappear at 
particular instants, so that one has a simple model of the creation/annihilation 
process.    

\begin{wrapfigure}[13]{r}{80mm} 
\vspace{-8mm}
\centering 
\includegraphics[width=80mm]{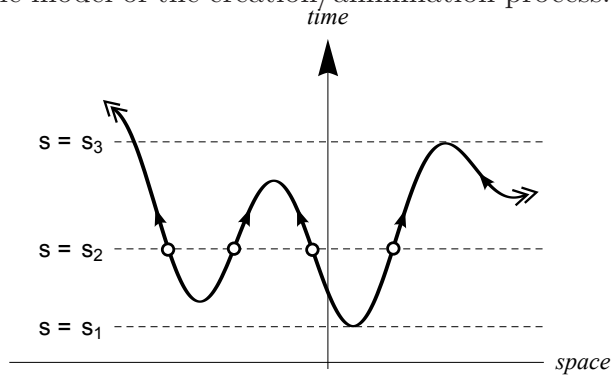}

\caption{Generic worldline, numerous pointlike ``particles'' (at $s=s_2$) and   
creation (at $s=s_1$) or annihilation (at $s=s_3$) events}
\label{img:figure1} 

\end{wrapfigure}

It should be noted that the Wheeler conjecture on the  
``one-electron Universe'' and, especially, on a ``positron as moving backwards 
in  time electron'' had been in fact explicitly initiated by the pioneer works of 
E.C.G. Stueckelberg~\cite{Stueckel1,Stueckel2}. He assumed the existence of 
worldlines of general type (forbidden in the canonical STR) that contain the 
segments corresponding to the superluminar 
velocities of particles' movements (\fref{img:figure1}). Then, the 
corresponding (hyper)plane 
of equal values  of the time-like coordinate $s=s_2$ intersects a worldline in a 
(generally, great) number of points. Physically, these form an 
ensemble of identical particles located on a single worldline. 

If, in the course of time, the coordinate $s$ is assumed to increase monotonically,  
then a pair of particles can appear at a particular instant $s=s_1$ 
(or disappear at $s=s_3$). These events model the processes of creation 
(annihilation) of a pair ``particle-antiparticle''.

Most likely, Stueckelberg~\cite{Stueckel1} himself considered $s$ as a 
{\it fourth coordinate} and did not assume it to be a real physical  
{\it evolution parameter}. As to the latter, he introduced a 
time-like parameter $\lambda$ which monotonically increases along the trajectory and  
is proportional to the proper time $\tau$ of a particle. After this, all equations of the 
theory can be formally represented in a relativistic invariant form. On the 
other hand, segments of the trajectory corresponding to the opposite increments 
of $\tau$ and $\lambda$, namely, to $ds/d\lambda<0$, were regarded as representing 
the backwards-in-time motion of an antiparticle. 

However, $\lambda$-parametrization is in fact parametrization for the history 
of an {\it individual} particle: this is in evident conflict with the concept of 
the ``one-electron Universe''. In order to preserve the ensemble of identical 
particles on a unique worldline, one should consider only $s$ as the ``true'' 
time. Then, however, velocities of ``particles'' should also be measured with 
respect to the parameter $s$ and are necessarily superluminar at some segments 
of their history (and even infinite at the annihilation points, see below). 
Stueckelberg himself fully comprehended this difficulty and wrote, 
in particular: ``Ceci, et d'autres consid\'erations d'ordre causal, nous 
semble \^etre in argument important contre l'hypoth\'ese de l'existence de 
telles forces ({\it that are responsible for redundant curvature of a worldline, 
resulting in superluminar velocities, {\bf V.K.,I.Kh.}}) 
malgr\'e la covariance de leur repr\'esentation''
~\cite[p.~592]{Stueckel1}. 

Subsequently, numerous approaches exploiting Stueckelberg's ideas 
(including his specific interpretation of the wavefunction, 
action functional and Lagrangian, etc.) came to be known as 
{\it parametrized relativistic theories} (see, e.g.,~\cite{Fanchi} 
and references therein). In most part of them, the additional time-like parameter 
had been treated as {\it a Lorentz invariant evolution parameter} or even as 
{\it absolute Newtonian time}~\cite{Enatsu,HorwPiron}~\footnote 
{Remarkably, in~\cite {Cawley} the invariance of Stueckelberg's action 
with respect to the {\it Galilean transformations} had been proved}. 
Nonetheless, the ultimate physical meaning of the variable $s$ is still 
unclear. Pav\u{s}ic~\cite{Pavsic} even considered it as an ``evolution parameter 
that marks an observer's subjective experience of {\it now}'' and tried 
to relate this to the process of localization of a particle's wave packet 
(to the collapse of wavefunction). One way or another, multiple ``particles'' 
on a single worldline related to {\it one and the same value of $s$}, are not 
causally connected and cannot be {\it simultaneously} detected by an observer. 

These and similar considerations reveal a lot of problems which arise 
under one's attempts to realize the ``one-electron Universe'' conjecture. 
However, the Stueckelberg-Wheeler-Feynman idea is too attractive to be 
abandoned at once. On account of the above-mentioned Galilei-invariance 
of Stueckelberg's construction, at the first step it seems quite natural to 
consider a purely non-relativistic picture of processes 
represented in \fref{img:figure1}~\footnote{Despite the generally accepted belief in 
the indissoluble connection of the annihilation/creation processes with 
relativistic structures}. {\bf The Galilean-Newtonian picture is the one that 
we accept in the main part of the paper}; it allows for a self-consistent 
realization of the ``one-electron Universe'' conjecture.  

Specifically, our main goal throughout the paper is 
to obtain {\it the correlated dynamics} of identical point-like particles 
from {\it the purely algebraic properties of a single worldline}~\cite{Abstr} and  
{\bf without any resort to the Lagrangian structure, differential 
equations of motion or other standard constituents 
of physics}. In this regard, our approach is 
quite different from and {\it much more radical} than those of Stueckelberg and 
Wheeler-Feynman.      
 
In order to realize the ``one electron Universe'' paradigm analytically, 
instead of the definition of a worldline in a habitual parametric form 
(and in the simplest parametrization $x_0 =s$)
\be{param}
x_a = f_a (s)~~~~~(a=1,2,3),  
\ee          
we define it (as is widely accepted for curves in mathematics) in an   
{\it implicit} form, i.e. through a system of three algebraic equations   
\be{implicit}
F_a (x_1,x_2,x_3,s) =0.
\ee

Then again, for any value of the time-like coordinate $s$, one generally 
has a whole set $(N)$ of real roots of this system, which define a 
correlated kinematics $x_a = f^{(k)}_a (s)$ of the ensemble of identical 
point-like singularities on a unique worldline~\footnote{In the STR, equations 
defining a worldline, at least in the simplest parametrization (\ref{param}), 
do not contain any trace of relativistic structure as well. It is therefore 
admissible to preserve the relativistic term ``worldline'' in the considered 
Galilean-Newtonian picture}.

It is worth noting that the 
{\it copies} arising via this algorithm (a l\'a Stueckelberg) {\it exist by 
themselves}. Their appearance is not related {\it \'a priori} to the existence 
of an ``observer'' or to the procedure of ``registration''. Thus, these 
identical particle-like formations  do not have direct connection 
either with the concept of duplicons~\cite{PIRT_dubl} or with the 
``charges-images'' of Bolotovskii~\cite{Bolot} mentioned above.

Multiple properties and ``events'' related to particle-like formations  
defined by the roots of system (\ref{implicit}) are considered in section 2 
and illustrated therein by a rather simple example. We restrict ourselves to  
{\it plane} motion  and to a {\it polynomial} form of {\it two}  
generating functions in ({\ref{implicit}). Particularly, we take into account 
not only real roots but complex conjugate roots as well: the latter turn out 
to have an independent physical sense and correspond to {\it another kind of 
particle-like formations}. 

In section 3, a short excursus into the methods of the 
mathematical investigation of the solutions of system (\ref{implicit}) 
(in the 2D case) of a generic polynomial type is undertaken. In the main, these 
methods make use of the so-called {\it resultants} of two polynomials. 
We demonstrate, in particular, that the {\it Vieta formulas} well known for 
a single polynomial equation,  naturally arise in the 2D case too. 
Quite remarkably (in the key section 4), the latter not only ensure the correlations between the  
positions and dynamics of different particles in the ensemble but also 
{\it reproduce the generic structure of Newtonian mechanics} and, 
in particular, lead to the satisfaction of the {\it law of momentum conservation}  
(in  the special inertial-like ``reference frames'')!

In the next section 4, we outline possible ways to appropriate 
{\it relativization} of the theory. In particular, we discuss the 
problem and possible advantages of the introduction of an external 
``observer'' into the scheme. Alternatively, we try to define the 
``second time'' parameter  
in the spirit of old conjecture of F. Klein et al. about the {\it universal 
light-like velocity of all of the matter pre-elements in the extended physical 
space} (4D in our case). This can be treated as a reformulation of the STR and  
could make the structure of the principle system ({\ref{implicit}) consistent 
with the {\it relativistic} mechanics.  

Section 5 contains concluding remarks on the motivations and actual 
developments of the presented scheme. As its important part, the paper also  
contains an appendix. Therein, the surprisingly rich dynamics defined 
by the simple polynomial system introduced in section 2 is retraced in detail, 
with the help of numerical calculations and graphical representations of the 
results. {\it One can also see an impressive animation of the dynamics in the 
supplementary data enclosed to the paper}. 

\section{Two kinds of point-like particles: algebraic kinematics}

Consider for simplicity the case of {\it plane motion}~\footnote
{We suspect that generalization to the physical 3D case will only be
technically more complicated but no problem of principal character will 
arise during it} and a curve  
defined implicitly through a system of two independent {\it polynomial} 
equations {\it with real coefficients}
\be{plane}
F_1 (x_1,x_2,s)=0,~~~F_2 (x_1,x_2,s)=0, 
\ee  
where $s\in {\mathbb R}$ is the particular coordinate which, in addition, 
plays the role of the {\it evolution parameter}: its variations  will be assumed 
{\it monotonic}. As was argued in the introduction, one can think of $s$ 
as being a Newtonian-like {\it absolute global time}. 

Restriction by a polynomial form of functions $F_1$ and $F_2$ is motivated by 
the fact that only in this case is one able to obtain the complete set of solutions 
to (\ref{plane}). Moreover, the roots are then explicitly linked via the Vieta 
formulas. This results in the {\it identical satisfaction of the law of momentum 
conservation} (see section 4 below).

It should be particularly emphasized that we consider the system (\ref{plane}) 
as the only one 
whose properties  we shall study throughout the paper: we do not intend to 
supplement it by any additional equations or statements of a physical or 
mathematical nature which do not explicitly follow from (\ref{plane}).

For any $s$, the system (\ref{plane}) generally has a finite ($N$) number of roots 
$\{x^k_1,x^k_2\},~~k=1,2,...N$. These define the positions of $N$ {\it 
identical point-like particles} at the instant $s$ on a 2D {\it trajectory curve} 
\be{traj}
F (x_1,x_2)=0,
\ee
whose form can be obtained from (\ref{plane}) after the elimination of $s$ and which, 
generally, {\it consists of a number of disconnected (on ${\bf R}^2$) components}. 
With monotonic growth of time $s$, particles move along the trajectory 
curve with arbitrary velocities, and their number is (almost always) preserved. 

However, at  particular discrete instants of $s$, say, at $s=s_0$, a pair of 
the real roots of (\ref{plane}) turns into one {\it multiple} root and then 
becomes a pair of {\it complex conjugate} roots. Consequently, the corresponding 
particles merge (collide) at $s=s_0$ at some point $\{x^0_1,x^0_2\}$ 
and then disappear from the real slice of space. Such an ``event'' can 
serve as a model of the {\it annihilation process}. Conversely, at  
another instant a  pair of the real roots can appear, modelling the process of 
{\it pair creation}. 

It should be noted nevertheless that one cannot ignore the formations which 
correspond to complex conjugate roots of (\ref{plane})  and ``live'' in the 
{\it complex extension} of real space. This fact will become evident in the next 
section, while at the moment we only remark that such formations can be 
depicted with respect to {\it equal real parts} of their coordinates. 

From this viewpoint, a pair of complex conjugate roots corresponds to a {\it 
composite} particle that consists of two 
parts coinciding at ${\bf R}^2$ but possessing opposite 
additional ``tails'' represented by imaginary parts of coordinates. 
For brevity, we shall call particle-like formations represented by the real roots of 
the system (\ref{plane}) ``R-particles'', and by complex conjugate pairs of 
roots ``C-particles''.
    
The condition for annihilation/creation events can be easily specified as that for 
{\it multiple} roots of the system (\ref{plane}) and has the form 
\be{multiple}
\det \Vert \frac{\prt F_A}{\prt x_B} \Vert = 0,~~~A,B,... =1,2.
\ee
Together with (\ref{plane}), condition (\ref{multiple}) defines a complete 
set of instants (and space locations) indicating  when (and where) such events 
do occur. 

We can now present a simple example of the issues exposed above. 
Let us take the functions $F_1$ and $F_2$ in (\ref{plane}), say, in the following 
({\it randomly selected}) form:
\begin{equation}\label{example} \left \{ 
\begin{array}{ll} F_1(x,y,s) = -2 x^3+y^3+sx+ sy+y+2 = 0, \\ 
F_2(x,y,s) = -x^3-2 x^2 y+s+3 = 0 \end{array} \right. 
\end{equation}                           

 \begin{wrapfigure}[18]{r}{0.5\textwidth}
\vspace{-5mm}
\centering 
\includegraphics[width=0.5\textwidth]{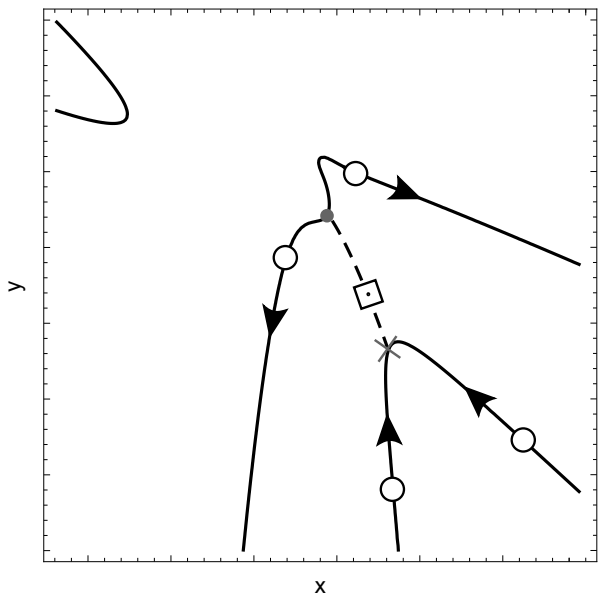}
\caption{Three branches of the trajectory of R-particles and the typical 
succession of events (annihilation - propagation of a C-particle - creation) 
}
\label{img:figure2}
\end{wrapfigure}

Eliminating $s$, one obtains the trajectory
 (on the real space slice) which  
consists of {\it three disconnected components} (\fref{img:figure2}).
Then, via elimination of $y$, one reduces the system (\ref{example}) 
to a single polynomial equation $P(x,s)=0$ 
of degree $N=9$ in $x$ and with coefficients depending on $s$.
The latter evidently allows for full analysis and numerical calculations. 
As a result, one obtains that, for any $s$, 
there exist precisely nine solutions of the system, some of them being real while others 
are complex conjugate. Analysing condition (\ref{multiple}) (or, equivalently, 
the structure of {\it discriminant} of the polynomial $P(x,s)$), one concludes 
that there are six ``events'' which correspond to the following (approximate) 
values of the global time $s$:
$-97.3689$; $-4.0246$; $-3$; $-2.7784$; $-2.7669$; $+2932.49$. Some of these relate to 
annihilation (merging) events  whereas others relate to creations of a pair.  
{\bf In the appendix (and in the animation file available as supplementary data) 
one can find  many details of the surprisingly rich dynamics}, including  the 
processes of annihilation of two R-particles 
accompanied by  the birth of a composite C-particle and vice versa. 
One also observes that the created ``C-quantum'' travels {\it between} two 
disconnected branches of the real trajectory and arrives at the second branch  
where gives rise to a divergent pair of real R-particles (creation of a pair), 
see also figure 2. Remarkably, this strongly resembles the process of 
{\it exchange of quanta} specific to QFT. 

Two peculiar aspects of the considered algebraic dynamics can be observed. 
The first one is the surprisingly great ``last'' critical value of the time 
parameter $s\approx 2932.49$,  despite the numerical coefficients in 
(\ref{example}) which all are of order 1. Thus, the ``history of a Universe'' 
defined via (\ref{example}) turns out to be unexpectedly long! It is not yet 
clear whether this property is of a particular or generic nature. 

The second aspect relates to impossibility to establish a unique parametrization 
$x_A =x_A(\lambda), \lambda \in {\mathbb R}$ for all the three disconnected 
branches of the trajectory. Here $\lambda$ is a parameter monotonically 
increasing along the trajectory and exploited, in particular, by 
Stueckelberg himself. From this impossibility it follows that distinction 
of particles from  antiparticles (say, ``electrons'' 
($ds/d\lambda >0$) from ``positrons'' ($ds/d\lambda <0$)) can be established 
{\it quite independently on each branch of the trajectory}. 
One can speculate whether this fact could be useful for the 
explanation of the particle/antiparticle asymmetry.   

To conclude, let us obtain the expression for  velocities of individual 
particles with respect to global time $s$. Introducing the canonical  
parametrization for an individual R-particle as $x_A = x_A (s)$ and then taking 
the total derivative with respect to $s$ (denoted by a ``dot'') in (\ref{plane}),  
one obtains 
\be{deriv}
0= \dot F_A + \frac{\prt F_A}{\prt x_B} \dot x_B, 
\ee  
whence it follows
\be{veloc}
\dot x_C = - R^A_C \dot F_A, 
\ee
where $R^A_C$ is the inverse matrix, 
\be{invers}
R^A_C~\frac{\prt F_A}{\prt x_B} = \delta^B_C.
\ee
Comparing (\ref{veloc}) with condition (\ref{multiple}), one concludes that 
at the instants of annihilation/creation, the velocities of both particles involved 
in the process are necessarily infinite~\footnote{This can also be seen just from 
\fref{img:figure1}, since at such instants one obviously has $ds=0, dx_A\ne 0$}.  
In the framework of the Galilean-Newtonian picture assumed throughout the paper, 
this property cannot cause any objection. Nonetheless, quite similar to 
the Stueckelberg's approach, 
at this point one encounters severe problems with causality and other    
principal statements of the STR. We consider these problems in section 5. 

\section{Resolving a system of polynomial equations: resultants and eliminants} 

Let us concentrate now on the procedure of resolution of the system of 
polynomial equations (\ref{plane}) of {\it generic type} 
(below we have made  obvious redesignations $x_1\mapsto x,~~x_2\mapsto y$), 
\begin{equation}\label{polynom} \fl \left \{ 
\begin{array}{ll}  F_1(x,y,s) = [a_{n,0}(s) x^n + a_{n-1,1}(s) x^{n-1}y+...+a_{0,n}(s) y^n]+...+ a_{0,0} (s)=0,\\ 
F_2(x,y,s)= [b_{m,0} (s) x^m +b_{m-1,1} (s) x^{m-1}y+... +b_{0,m}(s)y^m]+...+b_{0,0}(s) =0.\end{array} \right. 
\end{equation}
Only forms of the highest ($n$ and $m$, respectively) and the least orders are 
written out in (\ref{polynom}). Both polynomials are assumed to be functionally 
independent and irreducible, while all the coefficients $\{a_{i,j}(s),b_{i,j}(s)\}$ 
depend on the evolution parameter $s$ and take values in the field of real 
numbers $\mathbb R$. 

Rather surprisingly, not much is known about the properties of solutions to  
a nonlinear system of polynomial equations. Of course, some results can be taken 
from those for the one dimensional case. For example, it is easy to demonstrate 
that all the roots  $\{x_0,y_0\}$ of such a system are either real 
($x_0$ and $y_0$ both together) or both entering in complex conjugate pairs. 
However, even the problem of explicit determination 
of the full number of solutions of (\ref{polynom}) over $\mathbb C$ from, say, 
the properties of coefficients and degrees of the polynomials $F_1,F_2$ is far from 
being completely resolved (in contrast to the one dimensional case)~\cite{Gelfand}. 

In practical calculations, however, it is quite possible to determine 
this number and evaluate approximately all the roots of the system ({\ref{polynom}), 
for both real and complex conjugates. To do this, the most convenient method 
is perhaps the method of {\it resultants}~\cite{Gelfand,Morozov}. To be precise, 
let $\{x_0,y_0\}$ be a 
solution to (\ref{polynom}); then, for $y=y_0$ fixed, both equations  
({\ref{polynom}) on $x$ should have a {\it common root} $x=x_0$. 
The necessary and sufficient condition for this is well known: 
\be{resultY}
R_x(y) = g_N (s) y^N + g_{N-1} (s) y^{N-1} + ... +g_0 (s) =0
\ee
where $R[F_1(x),F_2(x),x] \equiv R_x(y)$ is the resultant of two polynomials  
$F_1,F_2$ via $x$ taken at the condition $y=y_0$ (for simplicity the index $0$ 
is omitted in (\ref{resultY}) and below). The structure of the resultant 
(which in this case is often called {\it eliminant}) is 
represented by the determinant of a {\it Sylvester matrix} 
(see, e.g.,~\cite{Morozov,Prasolov}).  
Coefficients $\{g_{I}(s)\}$ depend on $\{a_{i,j}(s),b_{i,j}(s)\}$.

Analogously, one can exchange the coordinates, and after the elimination of $y$ 
arrive at the dual condition
\be{resultX}    
R_y(x) = f_K (s) x^K + f_{K-1}(s) x^{K-1} + ... +f_0 (s)=0 .
\ee
When the coefficients $a_{n,0},a_{0,n},b_{m,0},b_{0,m}$ 
{\it are all nonzero}, the leading terms in (\ref{resultY}) and (\ref{resultX}) are,  
as a rule,~\footnote{Precisely, in the case when the numerical coefficient given by 
any of equal resultants $R[F^n_1(1,y),F^m_2(1,y),y]\equiv 
R[F^n_1(x,1),F^m_2(x,1),x]$ 
is {\it nonzero}, $F^n_1$ and $F^m_2$ being forms of the highest degrees 
($n$ and $m$, respectively), in (\ref{polynom})} 
of equal degree $K=N=mn$ (see, e.g.,~\cite{Dickenstein,Brill,Utyashev}). 
Then, all their $(N=mn)$ solutions over $\mathbb C$ can be numerically 
evaluated and put in correspondence with each other to obtain $N$ solutions 
$\{x_k(s),y_k(s)\}, k=1,2,..N$ of the initial system (\ref{polynom}). 
If some of the above four coefficients turn to zero, the number of solutions can 
be less than (or equal to) the maximal possible value $mn$. 
Nonetheless, in this case all the solutions can still be (approximately) 
obtained through seeking of eliminants, with the help of a 
computational software program.   

In order to illustrate the above presented procedure, consider the following 
system of equations  (closely related to the previous system (\ref{example}), 
see section 4 below):
\begin{equation}
\label{exampleC} \fl
\left \{ 
\begin{array}{ll}                                                      
F_1= -2x^3+y^3 +6s^2x^2+ 3sy^2 -(6s^4-s)x +(1+s+3s^2)y+\\~~~~~~~2s^6+s^2+s+2=0, \\
F_2 = -x^3-2x^2y +(3s^2-2s)x^2+4s^2xy - (3s^4 -4s^3)x -2s^4y+\\~~~~~~~s^6-2s^5+s+3=0.
\end{array} \right.
\end{equation}
Using the computer algebra system ``Mathematica 8'', we easily find 
the eliminant $R_x(y)$ and come to the equation
\be{resY} 
R_x(y) = 17y^9+153s y^8+... = 0.
\ee
Analogously, we obtain the dual condition
\be{resX} 
R_y (x) =   -17x^9+153s^2x^8+... =0.
\ee
The sets of nine solutions of equations (\ref{resY}) and ({\ref{resX}) can 
now be obtained and put in one-to-one correspondence with each other to 
form nine solutions of the system (\ref{exampleC}). For example, at $s=1$  
the system has one real solution $\{x\approx 2.3079,~y\approx-0.4848\}$ 
(defining the position of one R-particle) and four pairs of complex conjugate 
roots (corresponding to four C-particles).

\section{Vieta's formulas and the law of momentum conservation}

We are now ready to consider {\bf the most important issue of the present 
publication}, namely the correlations of different roots and the related particles' 
dynamics. These correlations follow from the {\it Vieta formulas}, which are 
well known for the case of a {\it single} polynomial equation. 

Specifically, as we have seen above, any solution of a system of two 
polynomial equations (\ref{polynom}) can be reduced to a pair of dual 
equations (\ref{resultY}) and (\ref{resultX}) for {\it eliminants} -- 
polynomials in one variable each ($x$ or $y$, respectively). Thus, we have 
demonstrated that the {\it Vieta's formulas naturally arise in the 2D case too}.

Below we 
consider the generic case when the degrees of both eliminants are equal, $K=N$. 
Then, the first and simplest of Vieta formulas (linear in roots) 
look as follows:                    
\begin{equation} \label{Viet}
\left \{ 
\begin{array}{ll} 
N X(s):=x_1(s)+x_2(s)+...x_N(s)=-f_{N-1}(s)/f_N(s),\\
N Y(s):=y_1(s)+y_2(s)+...y_N(s)=-g_{N-1}(s)/g_N(s).
\end{array} \right.
\end{equation}

Obviously, quantities $\{X(s),Y(s)\}$ can be regarded as coordinates of the 
{\it center of mass} of the closed system of $N$ identical (and, therefore, of 
equal masses $m_1=m_2=...=m_N$) point-like particles with coordinates represented 
by the roots $\{x_k(s),y_k(s)\}, k=1,2,..N$, of the system (\ref{polynom}) 
and varying in time $s$. 

An important fact here is that complex conjugate roots also enter the 
left-hand part of the conditions (\ref{Viet}) though their imaginary parts cancel 
and do not contribute to the center-of-mass coordinates. This observation 
makes it obvious that such roots cannot be regarded as ``unphysical''; in  
contrast, they should be treated as a {\bf second type of particle-like 
formations} (C-particles) which ``appear/disappear'' in the processes of 
creation/annihilation of real R-particles and ``move'' in the extension of 
space between the components of the trajectory of the latter. Only real parts of these 
complex conjugate roots contribute to the center-of-mass coordinates 
(and to total momentum, see below) and, on the other hand, can be visualized in 
the physical space. We have exemplified such a visualization in section 2. 
As to the imaginary parts of such roots, they could be responsible for 
internal phases and corresponding frequencies of C-particles~\cite{Mink,PIRT_09};  
however, their true meaning is vague at the present stage of consideration. 
Notice also that the {\it effective mass} 
of a C-particle is in fact twice as great as that of an R-particle since any 
C-particle is represented by a {\it pair} of complex conjugate roots (and thus 
by their equal real parts on the physical space slice).

The right-hand part of equations (\ref{Viet}) indicate that, generally, the center of mass 
of such closed ``mechanical'' system of the R- and C-particles does not, 
generally, move uniformly and 
rectilinearly. However, one can treat this contradiction with Newtonian mechanics 
as a manifestation of the 
{\it non-inertial nature} of the reference frame being chosen. One has 
therefore the right to perform a {\it coordinate transformation} to another frame 
which would model the inertial 
properties of matter (recall that we assume only one single ``worldline'' to exist 
which represents ``all particles in the Universe''). 

In fact, it is easier to just find the distinguished reference frame in which the 
{\it center of mass is at rest}. To do this, let us return  to the 
eliminants (\ref{resultY}),(\ref{resultX}) and get rid of the 
terms of the $(N-1)$th degree, setting 
\be{transf}
x = \tilde x  - ( N-1) f_{N-1}(s)/f_N(s),~~~y = \tilde y  - (N-1) g_{N-1}(s)/g_N(s).
\ee 
Now one can rewrite the system (\ref{polynom}) in the new variables as 
\be{polyN}
\tilde F_1 (\tilde x,\tilde y,s) =0,~~~\tilde F_2 (\tilde x,\tilde y,s) =0
\ee
and consider it as describing the same closed ``mechanical'' system of $N$ 
particles in the {\it center-of-mass reference frame}. Indeed, equations on 
the eliminants (\ref{resultY}),(\ref{resultX}) in the new variables take the form 
\be{resN} \fl
\tilde R_y(\bar x) = f_N (s) \tilde x^N + 0 + ... +\tilde f_0 (s)=0 ,~~ 
\tilde R_x(\bar y) = g_N (s) \tilde y^N + 0 + ... +\tilde g_0 (s)=0 
\ee
and, according to Vieta's formulas (\ref{Viet}), one obtains 
\be{VietT} \fl
N \tilde X(s):=\tilde x_1(s)+\tilde x_2(s)+..\tilde x_N(s)=0,~~~N \tilde Y(s):=
\tilde y_1(s)+\tilde y_2(s)+...\tilde y_N(s)=0.
\ee

Differentiating then (\ref{VietT}) with respect to the evolution parameter $s$ one 
obtains the law of conservation of the projections $P_x,P_y$ of total 
momentum for a closed  system of identical ``interacting'' particles defined 
by equations (\ref{polyN}):
\be{conserv} \fl
P_x:= \dot x_1(s)+\dot x_2(s)+...+\dot x_N(s)=0,~~~P_y:=
\dot y_1(s)+\dot y_2(s)+...+\dot y_N(s)=0  
\ee
(the sign ``tilde'' is omitted for simplicity).
 
If necessary, one can now transfer to another {\it inertial} reference frame  
using a {\it Galilei transformation}, say, $y\mapsto y,~~x\mapsto x-Vs,~~V=constant$ 
in which the center of mass will move  uniformly and rectilinearly 
with velocity $V$; specifically, one gets $X(s) =Vs,~~Y(s)=0$. 

Repeating now the procedure of differentiation, one obtains from (\ref{conserv}) a  
universal constraint on instantaneous accelerations of interacting 
identical particles:
\be{constr} \fl
\ddot x_1(s) +\ddot x_2(s)+...\ddot x_N(s)=0,~~~
\ddot y_1(s)+\ddot y_2(s)+... \ddot y_N(s)=0,  
\ee
which, for simplest case of a system of two particles, leads to   
Newton's third law together with definition of the forces of mutual interaction 
(provided the equal masses are set unit, $m_1=m_2=1$):
\be{force} \fl
f^{(21)}_x = m_1\ddot x_1,~f^{(12)}_x = m_2 \ddot x_2,~
f^{(21)}_y = m_1\ddot y_1,~f^{(12)}_y = m_2\ddot y_2;
\ee 
\be{3dNewton} \fl
a^{(1)}_x(s)+a^{(2)}_x(s)= f^{(21)}_x + f^{(12)}_x\equiv 0,~~~ 
a^{(1)}_y(s)+a^{(2)}_y(s) = f^{(21)}_y+f^{(12)}_y \equiv 0.                                                                       
\ee

Essentially, for two particles the whole system of Newton's mechanics 
may be completely recovered (though a concrete form of the forces' laws 
themselves is not fixed by the equations of the worldline (\ref{polyN})).

Consider now the case of three particles constituting a closed mechanical system. 
Then, in order to resolve the universal constraint on accelerations (\ref{constr}) 
(say, along $x$ and, analogously, along $y$)
\be{3part}  
a^{(1)}_x(s)+a^{(2)}_x(s)+ a^{(3)}_x(s)  =  0,
\ee
one may {\it introduce} the forces of mutual action and reaction
\be{forcedeterm} \fl 
a^{(1)}_x(s)  =  f^{(21)}_x + f^{(31)}_x,~~a^{(2)}_x(s)  =  f^{(32)}_x + f^{(12)}_x,~~
a^{(3)}_x(s)  =  f^{(13)}_x + f^{(23)}_x, 
\ee
which should then satisfy Newton's third law:
\be{3partNewton}
f^{(21)}_x + f^{(12)}_x \equiv 0, ~~f^{(13)}_x + f^{(31)}_x \equiv 0,~~
f^{(32)}_x + f^{(23)}_x \equiv 0.
\ee  

However,  system (\ref{forcedeterm}),(\ref{3partNewton}) cannot be uniquely 
resolved with respect to the forces of mutual action-reaction.  Of course, this 
fact is valid for any number of particles $N\ge 2$ and is of a general importance. 
In other words, {\it in a closed mechanical system it is principally impossible to uniquely 
determine the contributions of partial forces of action-reaction making use only   
of observations on accelerations of all the individual particles}! This fact 
(probably, not so widely known) can be regarded as an indication that, generally, 
the $N$-body problem should be from the beginning formulated at the 
language of {\it collective interactions}. In this connection, general constraints 
(\ref{constr}) represent the {\it weakened} form of Newton's third law: 
{\bf The sum of all resulting forces acting on all particles in a 
closed mechanical system is zero}.

Let us now return to illustrating the general construction presented above 
of the model of the 
``mechanical'' system consisted of $N=9$ point-like particles and defined by the 
equations of the worldline (\ref{exampleC}). Since the terms of degree $8=N-1$  
in the eliminants (\ref{resY}),(\ref{resX}) are nonzero and one of the 
corresponding  coefficients,  moreover, depends on the time parameter $s$ in a 
{\it nonlinear} way, the total 
momentum is not conserved. Thus, (\ref{exampleC}) represents the 
worldline in a non-inertial reference frame. In order to make a transition 
to the center-of-mass frame, one has to perform, according to (\ref{transf}), 
the transformation of coordinates of the form 
\be{convers}
x = \tilde x + s^2,~~y=\tilde y + s . 
\ee
In the new variables, eliminants (\ref{resY}),(\ref{resX}) take the form 
\be{resback}
R_x(y) = 17y^9+0+(35+33s)y^7+... = 0;~\\
R_y (x) =   -17x^9+0+(4s-4)x^7+... =0,
\ee
whereas the defining system (\ref{exampleC}) 
turns out to be exactly the system of equations (\ref{example}) 
(already examined in section 2). It is now not difficult to check that the 
total momentum of all nine particles defined by the latter is the same at every 
instant $s$ and, precisely, equal to zero. Thus, equations (\ref{example}) and 
(\ref{exampleC})  represent in fact the same ensemble of identical particles 
in the inertial center-of-mass reference frame and in a non-inertial one, 
respectively.   

It is worth noting that, besides the simplest 
linear Vieta formulas (\ref{Viet}), there exist other nonlinear ones,  
the highest of which, say, looks as follows:
\be{VietH} \fl
x_1(s)x_2(s)...x_N(s) = f_0(s)/f_N(s),~~~y_1(s)y_2(s)...y_N(s) = g_0(s)/g_N(s).
\ee
In principle, it is possible to find a transformation of coordinates 
that will nullify a number of terms in the eliminants; in this 
case one would have, apart from the center of mass and the related 
total momentum conservations, other combinations of roots 
(and their derivatives) which would 
preserve their values in time (``nonlinear integrals of motion in the 
framework of Newtonian mechanics''?). However, such transformations are 
implicit in nature (see, e.g.,~\cite{Prasolov,Chebotarev}), and it is very 
difficult (if possible) to find a transformed form of 
the defining system of equations as a whole. 
The problem certainly deserves a further consideration.  
    
To conclude the section, let us say some words about the {\it law of energy 
conservation}.  In the framework of nonrelativistic mechanics under 
consideration (and in contrast to the law of conservation of momentum), this 
law requires the {\it potential energy} to be taken into account. At present we 
are not aware whether the concrete form of the latter can be determined from 
the algebraic equations of the worldline alone. 

Nonetheless, there exist some hints that structure of the forces' laws 
can indeed be encoded in the general properties of the worldline. For 
instance, as far back as 1836, C.F. Gauss made an interesting observation on 
the roots $\{z_k\},~~k=1,2,...,N$, of a single polynomial equation $F(z)=0$ of a 
general form (see, e.g.,~\cite[ch.~1]{Prasolov}). These define a set of identical 
particles located at the corresponding points of the 
${\mathbb C}$-plane. Consider now any root $z_0$ of the {\it derivative} 
polynomial equation $F^\p(z)=0$ (which does not coincide with a (multiple) root of the 
initial equation). Then, it corresponds to a {\it libration point} (point of 
equilibrium) for the resultant field of {\it radial} forces produced by all the roots    
$\{z_k\}$, under the condition that {\it these forces be inversely 
proportional to the distance}, $f_k\propto 1/\vert z-z_k \vert$ (and effective 
``charges'' of the sources are all equal). Unfortunately, we were unable to 
find an analogue of this remarkable property in the 3D case. However, the  
example indicates that even in the 2D case (and in the 3D one as well) the roots 
of the {\it derivative} equations for eliminants (\ref{resultY}),(\ref{resultX}), 
namely, $R_x^\p(y)=0,~~R_y^\p(x)=0$, define in fact a {\it new (third) kind 
of particle-like formations} whose dynamics can be correlated with others in a
quite nontrivial way. We intend to consider this issue in a forthcoming 
publication.

\section{Remarks on relativization of the scheme}

It is now necessary (especially, in account of one's claims to offer 
an explanation of the annihilation/creation processes) to seek for possibilities 
of {\it relativization} of the theory. The formal way used for this purpose by 
Stueckelberg and his followers, as was demonstrated in the introduction,  
seems to be unsatisfactory, since it forbids the realization of the ``one-electron 
Universe'' conjecture. On the other hand, whether one regards the invariant 
parameter $s$ as a ``true'' time (with respect to which velocities of 
``particles'' on the worldline should be defined), then the scheme comes into 
irreconcilable conflict with the principles of STR (causality problems, 
tachyonic behavior). Besides, the very sense of the $s$-parameter 
and its relation to other ``times'' (coordinate time, proper time etc.)  
still remains vague. 

In order to remove contradictions with STR, as the first natural step 
one has to explicitly introduce into the scheme an {\it observer} and 
consider the process of {\it detection} of the (R- and C-) particles. Specifically, 
one must supplement the system of equations like (\ref{polynom}) (generalized 
to the 3D case) by the {\it retardation equation} 
\be{retard}
c^2(t-s)^2 = (x_o (t)-x)^2 + (y_o (t)-y))^2 + (z_o (t)-z))^2 .
\ee
Here the functions $\{x_o(t),y_o(t),z_o(t)\}$ define the worldline of an 
observer while  $\{x,y,z\}$ are the coordinates of the particles' unique 
worldline implicitly depending on $s$ via the system (\ref{polynom}). At this step, 
{\bf the fundamental constant -- velocity of light $c$ -- enters the theory for 
the first time}. Moreover,  
the introduction of the light cone equation (\ref{retard}) clarifies the 
meaning of $s$ as a {\it retarded time} parameter. Now, at any instant  
of the laboratory time $t$ the observer receives light-like signals 
from the whole set of particles located on a single worldline but at 
{\it distinct instants of the retarded time $s$}. Besides, this procedure 
opens a possibility of escaping the tachyonic behavior of particles at hand. Indeed, 
velocities fixed by the observer with respect to his proper time and to the 
retarded time defined by localizations of particles themselves can be quite 
different~\cite{VelocObsrv}. We remark that on a complexified space-time 
background, the corresponding procedure has been already exploited in the 
afore-mentioned theory dealing with the ensemble of duplicons
~\cite{YadPhys,PIRT_09} and will be considered in more detail elsewhere. 

Another possibility to overcome the superluminar velocities relates to the old 
conjecture of F. Klein~\cite{Klein}, Yu.B. Rumer~\cite{Rumer} et al. that any 
pre-element of matter always has in fact the same velocity, constant in 
absolute value  
(and equal to the speed of light in vacuum $c$) but in a {\it multidimensional 
extension of physical space}. In order to realize 
this idea in our scheme, one should 
consider the 4D Euclidean space ${\bf E}^4$ (with $s$ being the fourth 
coordinate) and introduce the following definition of the time increment $dt$:
\be{Euclid}
c^2 dt^2: = c^2 ds^2+dx^2+dy^2+dz^2 ,
\ee
which is equivalent to the above statement about the universal total velocity ($=c$),
\be{univers}
u^2+\vec v^2 =c^2, ~~~~(u:=c\frac{ds}{dt},~~\vec v:=\frac{d\vec r}{dt}, ~~~
\vec r:=\{x,y,z\}).
\ee

Introduction of the Euclidean structure, instead of the habitual Minkowski geometry, 
looks rather marginal. However, G. Montanus~\cite{Montanus} had demonstrated 
that the so-called {\it Euclidean relativity} could reproduce the main effects 
of the STR. On the other hand, I.A. Urusovskii in an interesting 
series of papers~\cite{Urus1,Urus2,Urus3} combined the postulate on universal total velocity 
(\ref{univers}) with the conjecture on {\it universal uniform rotation} of particles 
in the ``additional'' space dimensions (precisely, three in number in his scheme) 
around the circle of the radius equal to  their {\it Compton length}. These two 
statements have deep consequences and allow, in particular, for visual 
geometrical explanation of many relations of 
quantum theory (for this, see also~\cite{PIRT11}). As to the related group of 
transformations, Urusovskii demonstrated that this status can be preserved 
by the Lorentz group, so that his scheme had been called the ``6D treatment of 
Special Relativity''~\cite{Urus1}.

In the framework of the scheme presented here, the Montanus-Urusovskii's  
approach is interesting in two aspects. The first one is rather evident: 
velocities of the considered particles, with respect to the newly  
defined time interval $dt$, become bounded from above and, in particular, 
approach to maximal possible value $c$ near the annihilation points. 

The second aspect deals with relativization of the expression for momentum. 
From (\ref{Euclid}) it follows (as one usually has in the STR):
\be{timerel}
ds = dt \sqrt{1-v^2/c^2}, 
\ee
so that the previous Newtonian expression for momentum (\ref{conserv}) 
(with ``restored'' equal rest masses $m$)
\be{nonrelmom}
P_x = m\dot x = m\frac {dx}{ds},~~
P_y =m \dot y =m\frac{dy}{ds},~~
P_z =m \dot z =m\frac{dz}{ds}
\ee
takes now the well-known relativistic form
\be{relmom} \fl
P_x = mv_x / \sqrt{1-v^2/c^2},~~P_y = mv_y / \sqrt{1-v^2/c^2},~~
P_z = mv_z / \sqrt{1-v^2/c^2}.
\ee
Remarkably, the generating law of conservation of the center-of-mass 
position (\ref{VietT}) contains no differentiations and therefore preserves 
its ``non-relativistic'' form. 

We are not ready to discuss here all the consequences of the introduction 
of the Euclidean time increment  (\ref{Euclid}), the more so that some of them 
seem to differ from those required by the STR. It is only noteworthy that, 
geometrically, the corresponding time interval $\Delta t$ is equal to the path 
length (arc length of the trajectory curve) and can be calculated via 
explicit integration. 

In account of the existence of the second kind of particles related to the 
complex conjugate roots (C-particles), the definition of time increment 
(\ref{Euclid}) should be in fact generalized as follows:
\be{defcomplextime}
c^2 dt^2: = c^2 ds^2+(dx^2+d\xi^2)+(dy^2+d\eta^2)+(dz^2+d\zeta^2),  
\ee
where $\{d\xi,d\eta,d\zeta\}$ are the imaginary parts of increments of the 
corresponding complex coordinates. 

Finally, we note that the introduction of the time increment in the form 
(\ref{Euclid})  makes the time kinematically {\it irreversible}: 
any movement in the physical 3D or in an extended (real or complexified) space, 
by definition, gives rise to an increase of the time value, $dt>0$.

\section{Conclusion}

Stueckelberg-Wheeler-Feynman's conjecture about identical particles 
moving along a unique worldline looks attractive not only from the 
``philosophical'' viewpoint. It easily solves, say, the paradox that  
point-like particles can meet at some points of the physical 3D space 
(even for a 2D space the codimension of such an event is zero!). Moreover, 
the very condition that all such particles-copies belong to the same curve  
turns out to be a rigid restriction which requires a strongly correlated dynamics 
of these copies that reproduces in fact the process of physical interactions. 

We have demonstrated that any generic system of polynomial equations like 
(\ref{polynom}) completely defines a single ``worldline'' and an ensemble of 
identical point-like particles located on it. Their dynamics with respect to 
the evolution parameter $s$ reproduces (via Vieta's formulas) the generic  
structure of Newtonian mechanics. After the choice of a special (inertial) reference 
frame, the dynamics obeys the law of momentum conservation (for the closed 
system of two kinds, R- and C-, particle-like formations represented by 
real and complex conjugate roots, respectively). 
This looks as an important indication 
of the {\bf purely algebraic origins of the structure of (Galilean-Newtonian) 
mechanics and of physical interactions in general}.

Of course, many problems of principle character, including those of particle-
antiparticle asymmetry and transition to relativistic description still 
remain unsolved. Two approaches to relativization of the theory had been 
presented in the previous section. Another possibility is based on a 
conjecture on the {\it complex geometry} of (extended) spacetime that results 
in a number of intriguing consequences and natural connections with the Kerr-type 
solutions of the Einstein-Maxwell electrovacuum equations, with R. Penrose's 
{\it twistors} and models of {\it extended} particles.  For many decades, the 
conjecture has been elaborated by E.T. Newman and A.Ya. Burinskii et al. In 
particular, in~\cite{Newman1,Newman2} and~\cite{Burin1,Burin2}, remarkable 
particle-like and string-like structures related to a ``complex worldline'' and 
to the introduction of a ``complex time'' parameter were discovered. On the 
other hand, in~\cite{Burin3} the structure of {\it multiparticle Kerr-Schild 
solution} was obtained by making use of twistor methods and the Kerr theorem
~\cite{Penrose}. All these properties seem to have an explicit relation to 
our ``unique worldline dynamics'' and can open the way to algebraic 
construction of a {\it nonstationary ensemble} of the Kerr-like particles 
in interaction.

At present, it is noteworthy that any particle from the ensemble under consideration  
can be naturally endowed with equal (elementary) electric charge and produces 
an electromagnetic field of the Lienard-Wiechert type. It is especially 
interesting that this field undergoes an {\it amplification} at the points of 
merging (annihilation/creation) of a pair of particles, so that one has a 
nontrivial {\it caustic locus} which can be naturally regarded as a set of 
{\it quantum-like signals} perceived by an external observer~\cite{Bolot,YadPhys}.

As to the identification of the considered point-like formations (matter pre-elements) 
with real particles, at the present stage of investigation this, of course, seems 
premature. Moreover, physical particles could be detectable only at 
discrete instants of merging  of two or more pre-elements only when 
they emit a quantum-like signal. In this way one naturally comes to the concept 
of the {\it dimerous electron}~\cite{YadPhys,PIRT_09} which was found to be 
especially useful in the geometric explanation of the {\it quantum interference 
phenomena}.

Generally, at first one could make an attempt to find the reasons for the 
``attraction''  of different roots and, presumably, for their ability to form a 
sort of (stable) {\it cluster} which could really represent elementary particles, 
nuclei, etc. At present this still looks like a barely achievable dream, though 
the results obtained herein give essential support to the realization of the 
program.  

\section{Acknowledgement} 

The authors are greatly indebted to Professors E.T. Newman, B.P. Kosyakov, 
A.L. Krugly, J. Riscallah, A. Wipf and both referees for comprehensive and 
valuable comments,  to V.I. Zharikov and M.L. Fil'chenkov for thorough 
editing of the text.

\appendix
\setcounter{section}{1}
\section*{Appendix}   

Making use of the general procedure described in section 3, let us examine 
in detail the dynamics defined by the polynomial system of 
equations (\ref{example}), namely the following one:
\begin{equation}\label{exampleD} \left \{ 
\begin{array}{ll} F_1(x,y,s) = -2 x^3+y^3+sx+ sy+y+2 = 0, \\ 
F_2(x,y,s) = -x^3-2 x^2 y+s+3 = 0 \end{array} \right. 
\end{equation}

The trajectory curve of particle-like formations 
represented by {\it real} roots 
of this system following from the elimination of the evolution parameter $s$, 
is defined by the equation 
\be{trajex}
x^4+3x^3y+2x^2y^2-2x^3+y^3+3x+2y-2 =0  
\ee
and consists of three disconnected components (see \fref{img:figure2}). 

The full expressions for equations on {\it eliminants} $R_y(x)$ and $R_x(y)$ of 
the system (\ref{exampleD}) are as follows (compare with (\ref{resback})):
\begin{equation}\label{elimexample} \fl \left\{
\begin{array}{lll}
R_y (x) = -17x^9+(-4+4s)x^7+(3s+25)x^6 +(4s^2+12+16s)x^4 +\\
(-3s^2-18s-27)x^3 +27s+s^3+9s^2+27=0; \\
R_x(y) = 17y^9+(35+33s)y^7+(-6s+52)y^6+(15s^2+34s+19)y^5 +\\
(40+8s-16s^2)y^4 +(49s+11s^2-s^3+113)y^3+(-50s-12-18s^3-72s^2)y^2 +\\
(148s^2+28s^3+208s+48)y - 64+s^4-48s^2-5s^3-96s=0.
\end{array}\right.
\end{equation}
One obtains from (\ref{elimexample}) that at any instant $s$ the system (\ref{exampleD}) 
has nine solutions, some of them composed of complex conjugate pairs; besides, 
since the terms of eighth degree are absent, the total momentum of the two 
types (R- and C-) of particles represented by real and complex conjugate roots is 
permanently equal to zero (the centre of mass reference frame). 
Values of $s$ that determine singular points for the solutions of (\ref{example}) 
related to the 
annihilation/creation events correspond to {\it multiple roots}~\footnote
{In order to determine these, one could use the explicit condition 
(\ref{multiple}). We, however, prefer another, more visual, method of  
discriminants}  of the equations 
(\ref{elimexample}), or {\it common roots} of the two systems of equations 
\be{singX}\left\{
\begin{array}{ll}
R_y(x)=0,\\
R_y^\p(x)=0,
\end{array}
\right.
\ee
and 
\be{singY}\left\{
\begin{array}{ll}
R_x(y)=0,\\
R_x^\p(y)=0,     
\end{array}\right.
\ee
where the ``prime'' denotes differentiation with respect to $x$ or $y$, respectively. 

Computing now {\it resultants} of the two polynomials in (\ref{singX}) or (\ref{singY}), 
which are in fact the so-called {\it discriminants} of equations (\ref{elimexample})  
one verifies that these two {\it have common factors}, so that 
critical values of parameter $s$ are obtained from the {\it real roots} of the 
equation
\be{critic} \fl
\begin{array}{lllll}
R_{common}(s)=(s+3)^3(-1030738720704832-2585288646749952s-\\~~~~~~~~~~~~~~2876632663642944s^2
-3915728526452064s^3-6758379899262912s^4-\\~~~~~~~~~~~~~~7627803495311328s^5
-5242401840993563s^6 -2294579103345501s^7-\\~~~~~~~~~~~~~~652002779260446s^8
-117671742918602s^9-12435143753367s^{10}-\\~~~~~~~~~~~~~~617360791689s^{11}
-4976985600s^{12} +1769472s^{13})=0.
\end{array}
\ee
Equation (\ref{critic}) has an obvious root $s=-3$ of multiplicity 3, and 
13 other roots of which only 5 turn out to be real. Thus, the system (\ref{elimexample})  
defines six critical values of the parameter $s$ at which some merging of roots and 
related particle-like formations take place. Approximate critical 
values of $s$ have been written out in section 2 and will be 
reproduced below. 
Corresponding coordinates of the points of merging are then readily obtained 
from the eliminants' equations (\ref{elimexample}).

Consider now a graphical representation of the successive dynamics of 
roots of the system (\ref{exampleD}) at different values of the time parameter $s$. 
To begin with, let us agree about the notation on figures. Three disconnected 
branches of the trajectory (\ref{trajex}) are denoted as A,B,C ones. 
Circles designate 
the positions of real roots (particles of the type R), and squares {\it real parts} of 
complex conjugate roots (particles of the type C), which are assumed thus 
to be located both in one and the same space point.  Arrows designate the 
direction of motion of roots under the positive increment of the parameter $s$. 
The roots are numbered in order to follow their successive dynamics and 
transmutations. By a grey cross or circle with corresponding inscriptions 
$s_k,~~k=1,2,..$, one denotes the positions and instants of the annihilation or 
creation events, respectively. Finally, by dotted lines some segments 
of the projection  of trajectories of complex conjugate roots onto the real 
plane are denoted, for a visual representation of the dynamics of the corresponding 
C-particles.

\begin{figure}[h]
        \centering
        \begin{subfigure}[b]{0.5\textwidth}
                \centering
                \includegraphics[width=\textwidth]{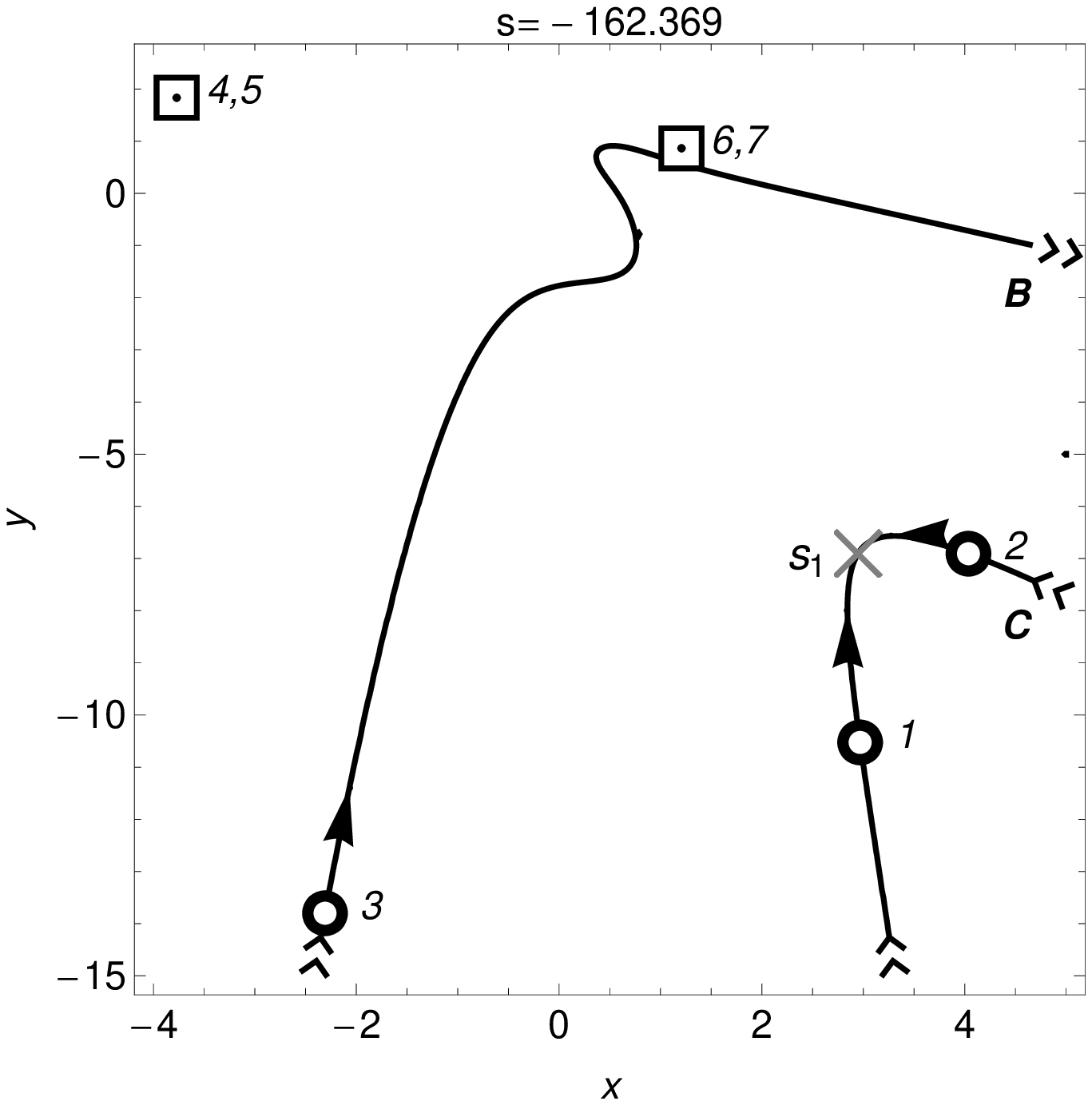}
                \caption{}
                \label{img:figure3}
        \end{subfigure}%
        ~ 
        \begin{subfigure}[b]{0.5\textwidth}
                \centering
                \includegraphics[width=\textwidth]{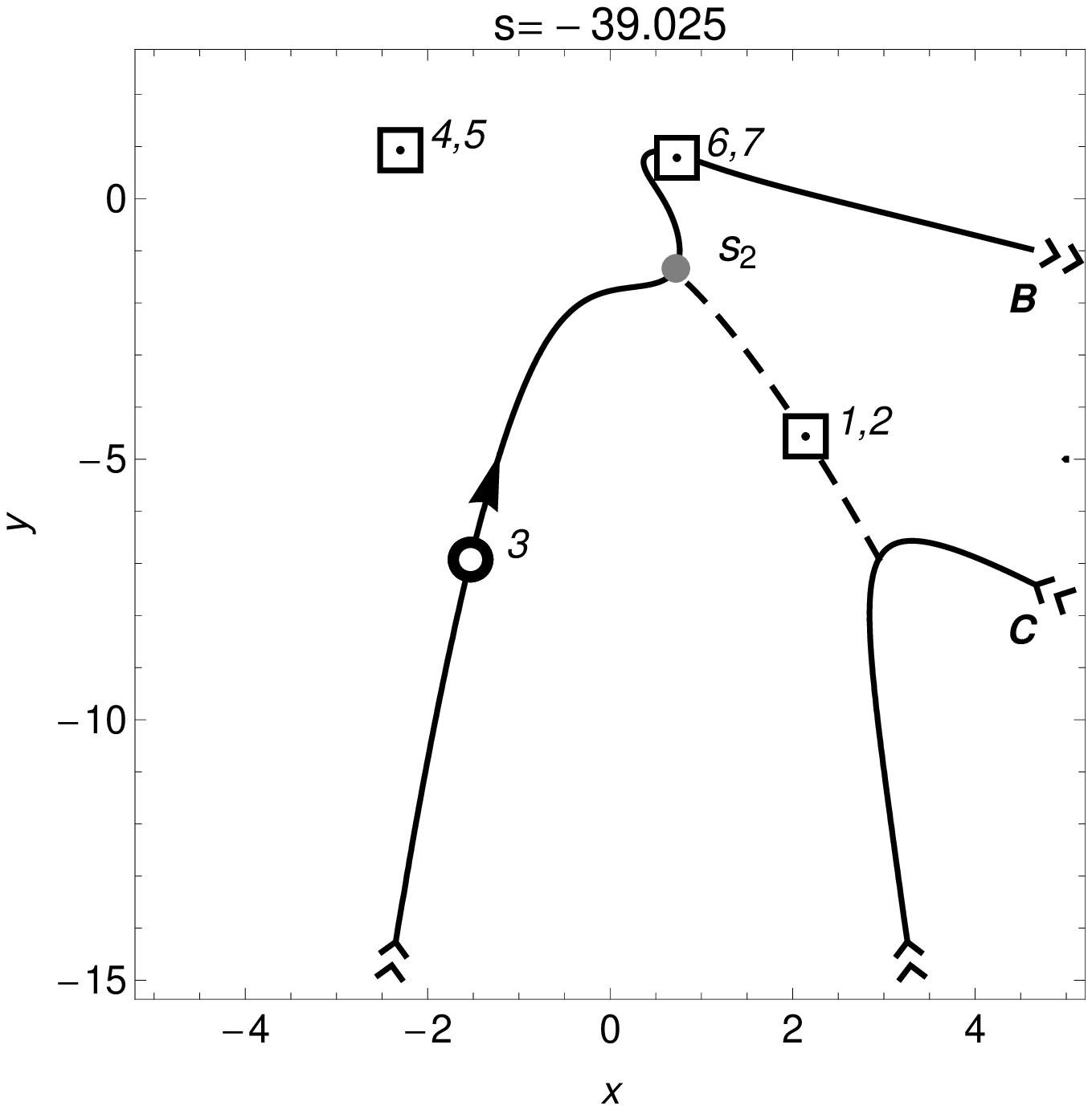}
                \caption{}
                \label{img:figure4}
        \end{subfigure}
        \vspace{-5mm}
        \caption{Disposition of roots of the system (\ref{elimexample}): 
(a) at $s\approx -162.37$; (b)~at~$s\approx -39.025$ (after first annihilation).}
				\label{fig:imgs}
				\vspace{-5mm}
\end{figure}

In \fref{img:figure3} one sees  that the real roots 1 and 2 move towards one 
another along the first branch of the trajectory C (\ref{traj}), up to their 
annihilation  at $s_{1}\approx -97.3689$.

\Fref{img:figure4} represents the intervening situation, when the above roots become 
complex conjugate and are under transition to the other branch B 
where they are expected to give rise to a new pair of R-particles, at 
$s_2\approx -4.025$. Note that one pair of complex conjugate roots 
is off the depicted space at \fref{img:figure3} and \fref{img:figure4} so that 
only seven roots are represented therein.

\begin{figure}[h]
        \centering
        \begin{subfigure}[b]{0.5\textwidth}
                \centering
                \includegraphics[width=\textwidth]{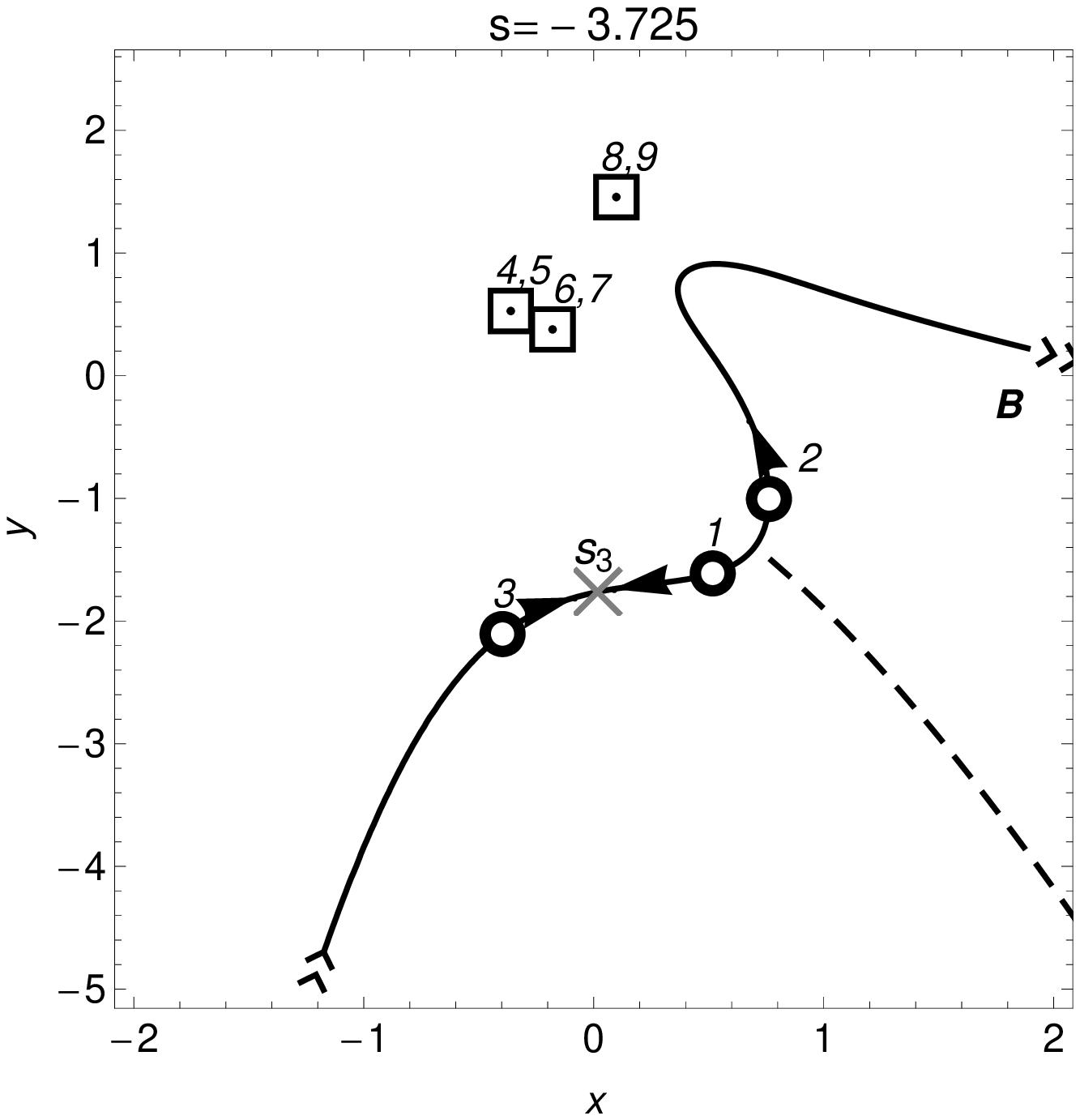}
                \caption{}
                \label{img:figure5}
        \end{subfigure}%
        ~
        \begin{subfigure}[b]{0.5\textwidth}
                \centering
                \includegraphics[width=\textwidth]{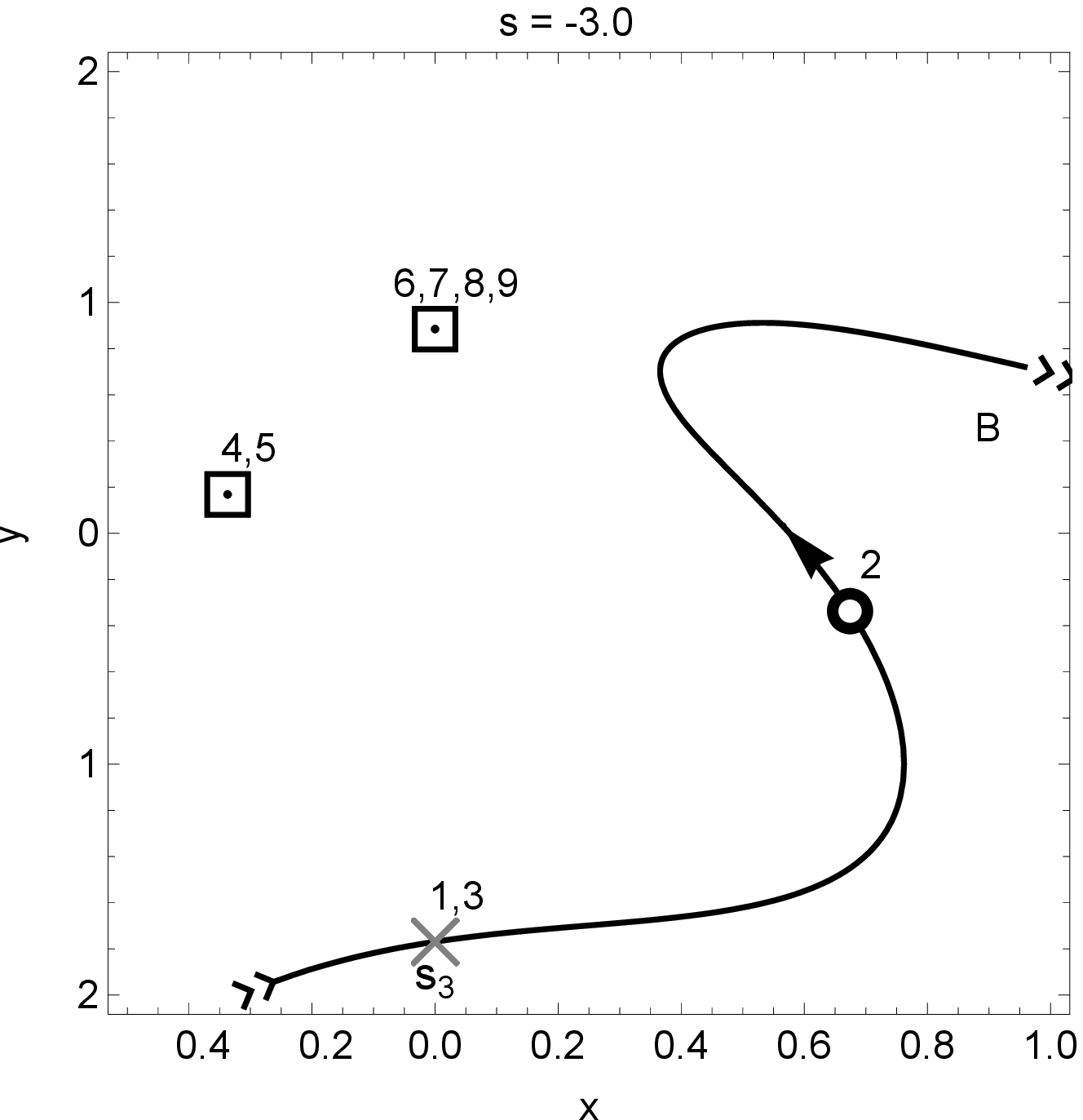}
                \caption{}
                \label{img:figure6}
        \end{subfigure}
        \vspace{-5mm}
        \caption{Disposition of roots of the system (\ref{elimexample}): (a) at 
$s\approx -3.725$ (after first pair creation); (b) at $s=-3$ (double merging).}\label{fig:imgs2}
        \vspace{-5mm}
\end{figure}

In \fref{img:figure5}, one sees that the considered roots 1 and 2 give rise to 
a pair of real R-particles (1 and 2) at the branch B of the trajectory. 
The root 3 moves towards  real root (1) and will 
merge with the latter at $s_3=-3$. Note that the third pair of complex 
conjugate  roots (8 and 9) appears in the space of vision so that 
the full number of roots (N=9) is depicted here and in the subsequent figures.

In \fref{img:figure6}, a peculiar situation of {\it double merging} is presented at $s_3=-3$ 
(recall that this is the exceptional root of multiplicity 3 of the equation 
for ``events'' (\ref{critic}). 
At this instant, besides the annihilation of 
two real R-particles (1 and 3) one has the merging of two complex 
conjugate pairs of roots (6,7 and 8,9) which {\it takes place 
in the space exterior to the real trajectory} (i.e. in the complex 
extension of the ``physical'' 3D space). 
\begin{figure}[h]
        \centering
        \vspace{-5mm}
        \begin{subfigure}[b]{0.5\textwidth}
                \centering
                \includegraphics[width=\textwidth]{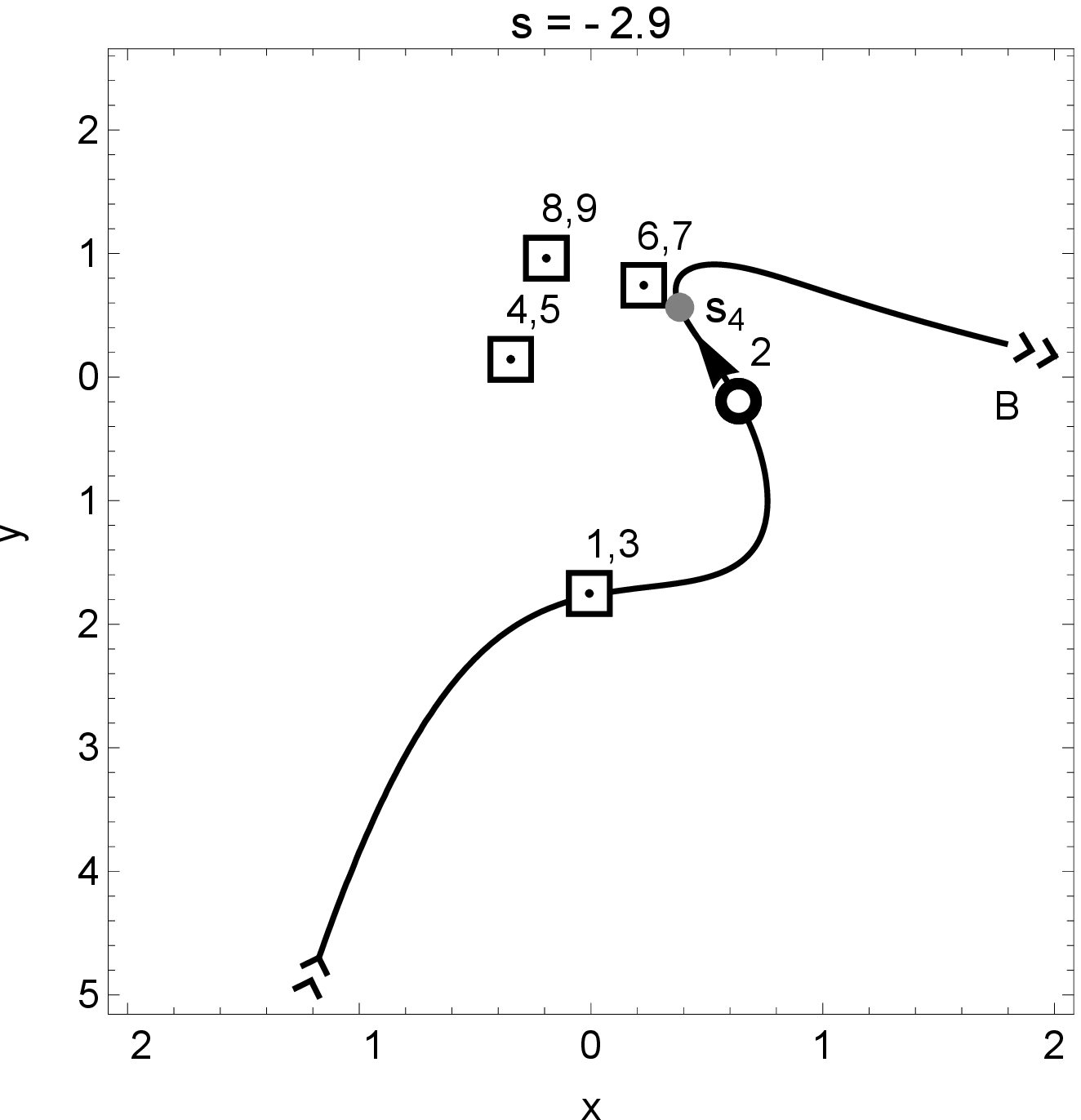}
                \caption{}
                \label{img:figure7}
        \end{subfigure}%
        ~
        \begin{subfigure}[b]{0.5\textwidth}
                \centering
                \includegraphics[width=\textwidth]{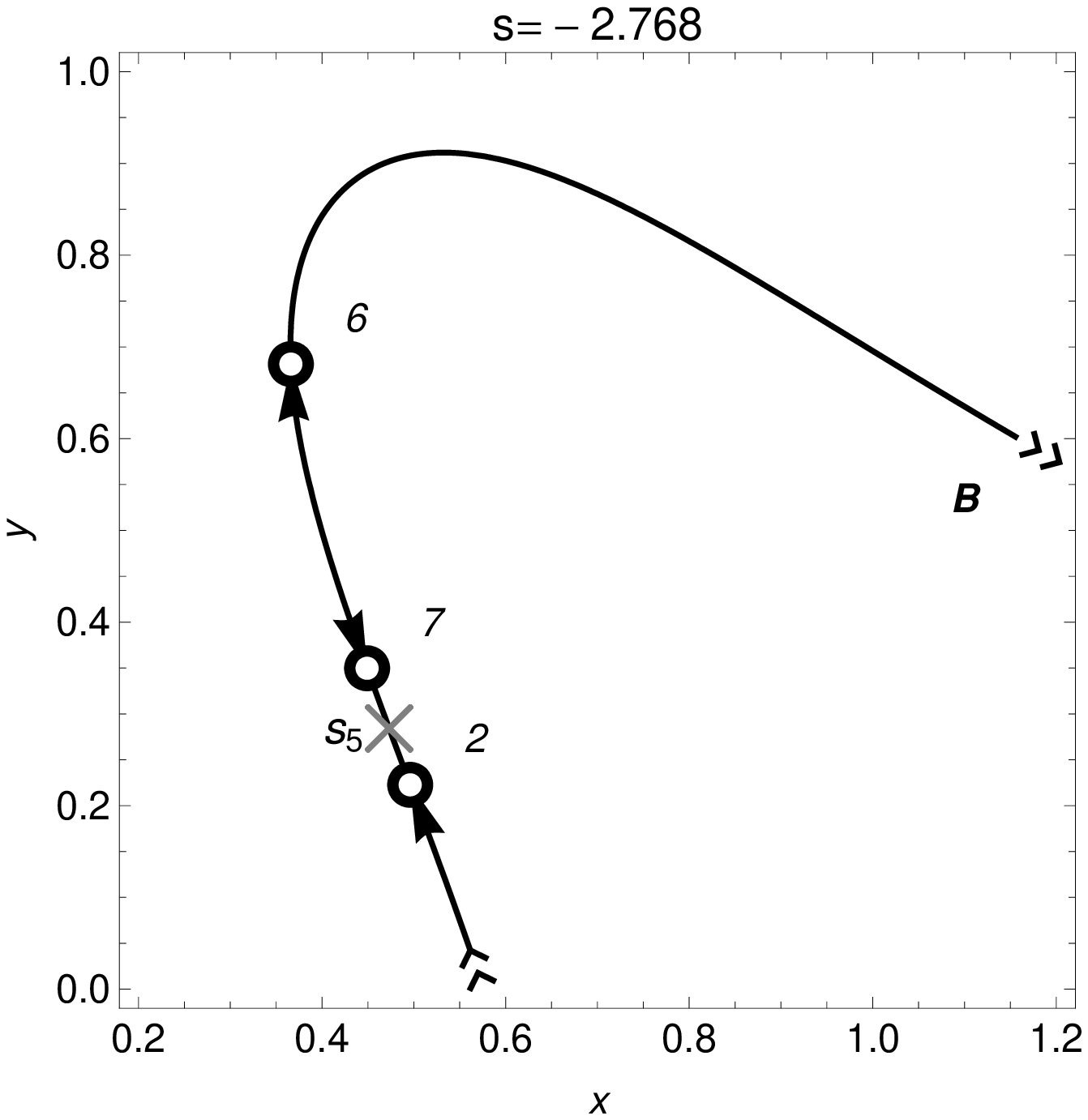}
                \caption{}
                \label{img:figure8}
        \end{subfigure}
        \vspace{-10mm}
        \caption{Disposition of roots of the system (\ref{elimexample}): 
(a) at $s\approx -2.9$; (b) at $s\approx -2.768$ (second pair creation).}
     		\label{fig:imgs3}
     		\vspace{-5mm}
\end{figure} 
In contrast to the merging of real 
particles, such an event {\it is not accompanied by annihilation of a pair}:
in what follows, the merged pairs deviate from one another, without any 
modification of their structure (see \fref{img:figure7}). 

In \fref{img:figure7}, one observes only one real root (2) while one pair of complex 
conjugate roots (6 and 7), after divergence with the other pair (8 and 9), 
moves towards  branch B of the trajectory where it will give rise to 
a pair of real roots (6 and 7) at the next moment $s_4\approx -2.78$. 

In \fref{img:figure8}, the two created real particles (6 and 7) move in opposite 
directions along the branch B of the trajectory. At the next moment, 
annihilation ot roots (2 and 7) at $s_5\approx-2.77$ is expected. The pair 
of complex conjugate roots moves towards the third branch A of the 
trajectory (to be seen at the next figure) which at the moment 
is still ``empty''. 

\begin{wrapfigure}[16]{r}{0.5\textwidth}
\vspace{-10mm}
\centering
\includegraphics[width=0.5\textwidth]{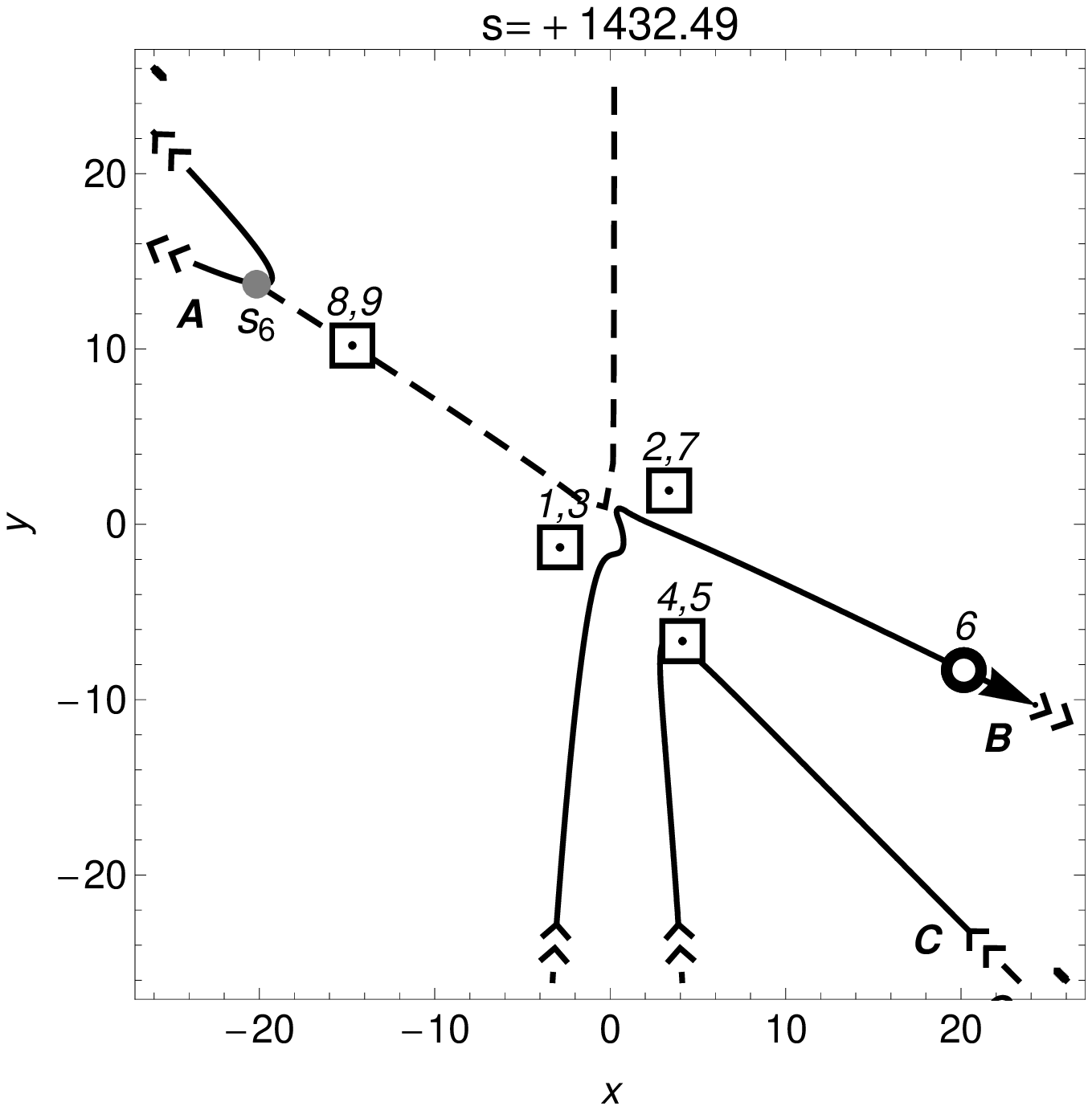}
\vspace{-5mm}
\caption{Disposition of roots of the system (\ref{elimexample}) at $s\approx 1432.49$ (after third annihilation).}
\label{img:figure9}
\vspace{5mm}
\end{wrapfigure}

In \fref{img:figure9}, the disposition of roots are presented 
at a much greater scale. After the 
annihilation of  roots 2 and 7 only one real R-particle (6) survives  
on the branch B. The pair of roots 8 and 9 moves (precisely, in 
complex extension of space) towards the third, ``empty'' branch of the 
trajectory A where the third pair creation is expected at the future 
moment $s\approx 2932.49$.  After this last event, there exist two real 
particles at branch A, one real particle at branch B and three pairs 
of complex conjugate roots (three C-particles). From now on, no other merging 
events do exist: the dynamics is in fact over.

\begin{figure}[H]
\vspace{5mm}
\centering
\includegraphics[width=0.6\textwidth]{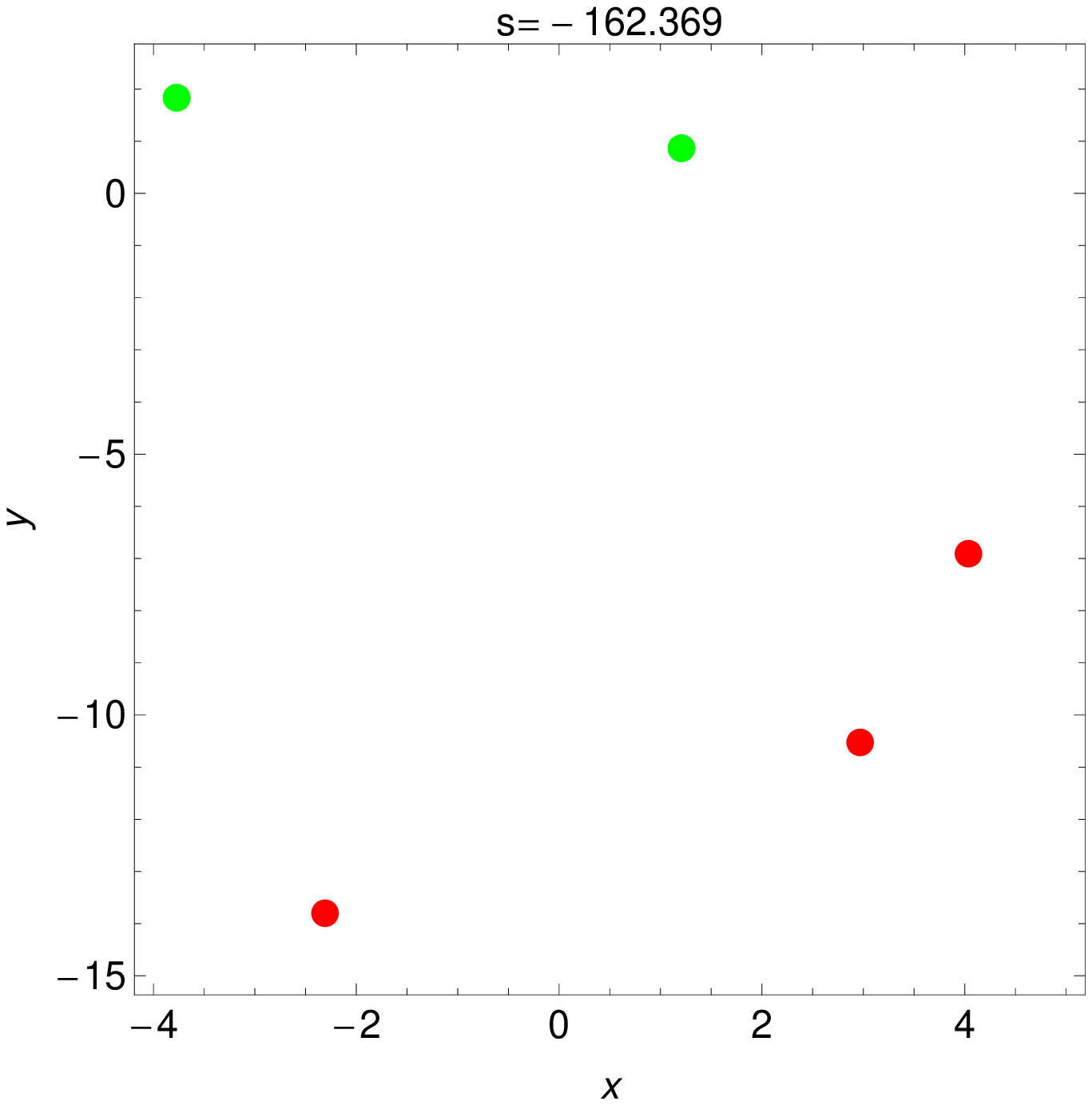}
\vspace{5mm}
\caption{A representative frame from the supplementary animation file animation.avi 
(1�898�036 bytes).}
\label{img:figure10}
\end{figure}   

Full animation of the above presented dynamics is accessible with the help  
of the enclosed file ``animation.avi'' (see \fref{img:figure10}). Note that, for better perception,  
the temporal and spatial scales are made variable throughout the presentation.

\end{document}